\documentclass[aps,superscriptaddress,showpacs]{revtex4}
\usepackage{graphicx}% Include figure files
\usepackage{dcolumn}% Align table columns on decimal point
\usepackage{bm}% bold math

\begin{document}

\title{Extrema bounds for the soft Pomeron intercept}
% Force line breaks with \\

\author{E. G. S. Luna}
\email{luna@ifi.unicamp.br}
\author{M. J. Menon}
\email{menon@ifi.unicamp.br}
\affiliation{
Instituto de F\'{\i}sica ``Gleb Wataghin''\\
Universidade Estadual de Campinas, UNICAMP\\
13083-970, Campinas, SP, Brazil
}

\date{\today}% It is always \today, today,
             %  but any date may be explicitly specified

\begin{abstract}
By using an extended Regge parametrization and taking into account the discrepancies 
in the 
high-energy $pp$ and $\bar{p}p$ total cross section data in  
both accelerator and 
cosmic-ray regions, we estimate extrema bounds for the soft Pomeron intercept. 
First we consider two ensembles of data with either the CDF or the 
E710 and E811 results for $\sigma_{tot}^{\bar{p}p}$ at 1.8 TeV, from which we 
obtain the bounds 1.102 and
1.081, respectively. These ensembles are then combined with the highest and lowest
estimations for $\sigma_{tot}^{pp}$ from cosmic-ray experiments (6-40 TeV), leading
to the upper and lower bounds 1.109 and 1.082, respectively. The effects of 
simultaneous fits to $\sigma_{tot}$ and $\rho$, individual fits to $\sigma_{tot}$,
and the influence of the subtraction constant in the dispersion relations are
also presented. Our global results favor the E710 and E811 data.
\end{abstract}

\maketitle

\section{\label{sec:level1}Introduction}

Analytic models for hadron-hadron scattering are characterized by simple 
parametrizations for the forward amplitude $F$ and the use of dispersion relation 
techniques to study the total cross section $\sigma_{tot}$ and the $\rho$ parameter 
(the ratio of the real to the imaginary part of the amplitude),

\begin{eqnarray}
\sigma_{tot}(s) = \frac{\textrm{Im}\ F(s,t=0)}{s},
\quad
\rho = \frac{\textrm{Re}\ F(s,t=0)}{\textnormal{Im}\ F(s,t=0)},
\label{sigrho}
\end{eqnarray}
where $t$ is the four-momentum transfer squared and $s$ the center-of-mass energy
squared.

In a recent work several aspects concerning the application of the analytic models
to $pp$ and $\bar{p}p$ elastic scattering have been studied \cite{alm03}. In particular, 
in the case of the Donnachie-Landshoff parametrization, investigation of discrepant 
estimations for the total cross sections from 
cosmic-ray experiments allowed to infer an upper bound for the soft Pomeron intercept,
namely, $1 + \epsilon = 1.094$. In addition, the effects of global vs individual fits to
$\sigma_{tot}$ and $\rho$, and the effects of the subtraction constant in the 
dispersion relations have also been analyzed and discussed.

In this report we extend the previous analysis in several ways, with focus on new
upper/lower bounds  for the soft Pomeron intercept: (1) we investigate the effect
of discrepant values for $\sigma_{tot}$ from accelerator experiments at $1.8$ TeV,
by selecting different ensembles of data that include either the highest (CDF) or
the lowest (E710/E811) results; (2) these ensembles are then combined with discrepant
estimations for $\sigma_{tot}$ from cosmic-ray experiments, 
now using as the lowest estimations the results
by Block, Halzen, and Stanev; (3) we use here, as a framework,
an extended parametrization with 
non-degenerate $C = + 1$ and $C = - 1$ meson trajectories.
As in the previous analysis we also present the effects of individual and global
fits to $\sigma_{tot}$ and $\rho$, and the effect of the subtraction constant.
Since the soft Pomeron exchange dominates the high energy behavior of the
total cross sections and the $pp$ and $\bar{p}p$ scattering correspond to the
highest energy interval with available data (including information from
cosmic-ray experiments), we shall limit our analysis to these processes.

The paper is organized as follows.
In Sec. II the essential formula of the analytic approach with the extended
Regge parametrization are presented. In Sec. III the discrepancies at both accelerator
and cosmic-ray domains are reviewed, and it the fit results through
four different ensembles of experimental information are presented. The conclusions and 
some final remarks are the contents of Sec. IV.

\section{Extended Regge parametrization}

The forward effective Regge amplitude introduced by Donnachie and Landshoff
has two contributions, one from a single Pomeron and the other from
secondary Reggeons exchanges\cite{dl92}.
The model assumes degeneracies between the secondary reggeons, imposing a common 
intercept for the $C=+1$ ($a_2, f_2$) and the $C=-1$ ($\omega,\rho$)
trajectories. This was the parametrization adopted in the previous paper 
\cite{alm03}.

Although the original fits by Donnachie and Landshoff have been performed 
only to the $\sigma_{tot}$ data, more recent analysis, treating global fits to
$\sigma_{tot}$ and $\rho$, have indicated that the best results are obtained with 
non-degenerate meson trajectories \cite{ckk,cmg}. In this case the forward
scattering amplitude is decomposed into three reggeon exchanges,
$
F(s) = F_{\tt I\!P}(s) + F_{a_2/f_2}(s) + 
\tau F_{\omega/\rho}(s), 
$
where the first term represents the exchange of a single Pomeron, the other
two the secondary Reggeons and $\tau = + 1$ ($- 1$) for $pp$ ($\bar{p}p$)
amplitudes. Using the notation $\alpha_{\tt I\!P}(0) 
= 1+\epsilon$, $\alpha_{+}(0) = 1 -\eta_{+}$ and $\alpha_{-}(0) = 1 -\eta_{-}$ 
for the intercepts of the Pomeron and the $C=+1$ and $C=-1$ 
trajectories, respectively, the total cross sections, Eq. (1), for $pp$ and $\bar{p}p$ 
interactions are 
written as
%\begin{subequations}
\begin{eqnarray}
\sigma_{tot}(s) = X s^{\epsilon} + Y_{+}\, s^{-\eta_{+}} + \tau Y_{-}\, 
s^{-\eta_{-}}.
\end{eqnarray}

The connection with the $\rho$ parameter is obtained by means of dispersion relations
and for the above parametrization
convergence is ensured by using analyticity relations with one subtraction. 
Defining $2 F_{\pm} \equiv F_{pp}\pm F_{\bar{p}p}$ these
relations read \cite{alm03}
 
\begin{eqnarray}
\nonumber
\textrm{Re}\ F_{+}(s) = K + s \tan \left[ 
\frac{\pi}{2} \frac{d}{d\ln s} \right] \frac{\textrm{Im}\ F_{+}(s)}{s}, \qquad
\textrm{Re}\ F_{-}(s) = \tan \left[ \frac{\pi}{2} \frac{d}{d\ln s} \right] 
\textrm{Im}\ F_{-}(s) , 
\nonumber
\end{eqnarray}
where $K$ is the 
subtraction constant. Within this formalism, Eqs. (1), (2) and (3)
lead to the following connection between $\rho(s)$ and $\sigma_{tot}(s)$:

\begin{eqnarray}
\nonumber
\rho(s)\ \sigma_{tot}(s)= \frac{K}{s} + X\, s^{\epsilon} \tan \left( 
\frac{\pi \epsilon}{2} \right) - 
Y_{+}\, s^{-\eta_{+}} \tan \left( \frac{\pi \eta_{+}}{2}  \right) 
 + \tau 
Y_{-}\, s^{-\eta_{-}} \cot \left( \frac{\pi \eta_{-}}{2} \right). 
\end{eqnarray}

\section{Discrepancies, strategies and fitting results}

The experimental information on $pp$ and $\bar{p}p$ total cross 
sections at the highest energies are characterized by discrepant results.
As is well known, in the accelerator region, the conflit concerns the results
for $\sigma_{tot}^{\bar{p}p}$ at $\sqrt s = 1.8$ TeV reported by the  
CDF Collaboration \cite{CDF} and those reported by the E710 \cite{E710} and the  
E811 \cite{E811a,E811} Collaborations (Fig. 1).
In the cosmic-ray region, $6\ \textrm{TeV} < \sqrt s \leq 40$ TeV, the discrepancies are
due to both experimental and theoretical uncertainties in the determination of
$\sigma_{tot}^{pp}$ from p-air cross sections. The situation has been recently 
reviewed in detail in
our previous paper \cite{alm03}, where a complete list of references, numerical results and
discussions are presented. As showed there, the highest predictions for $\sigma_{tot}^{pp}$
concern the result by Gaisser, Sukhatme, and Yodh \cite{gsy} together with those by
Nikolaev \cite{niko}. In the other extreme, the lowest values come from the results by
Block, Halzen, and Stanev \cite{bhs}. These extrema estimations
are displayed in Fig. 1 
(numerical values may be found in ref. \cite{alm03}).

Although, in principle, all available data in the accelerator region
could be used, it should be stressed that the difference between the CDF and the E710/E811 
results involves two standard deviations \cite{E811a}. This strong disagreement certainly 
indicates
the possibility of distinct scenarios for the rise of the total cross section and 
consequently
for the value of the Pomeron intercept. Moreover, despite the large error bars in the
extracted values of $\sigma_{tot}^{pp}$ from cosmic-ray experiments, the discrepancies
also presented can corroborate the distinction between the different scenarios. 
It is expected that answers to these questions will be provided by the new
values for $\sigma_{tot}$ and $\rho$ coming from the BNL RHIC, the Fermilab Tevatron-run 
II and the CERN LHC.

\begin{figure*}
\vspace{1.0cm}
\vglue -1.0cm
\hglue -8.5cm
\includegraphics[height=.30\textheight]{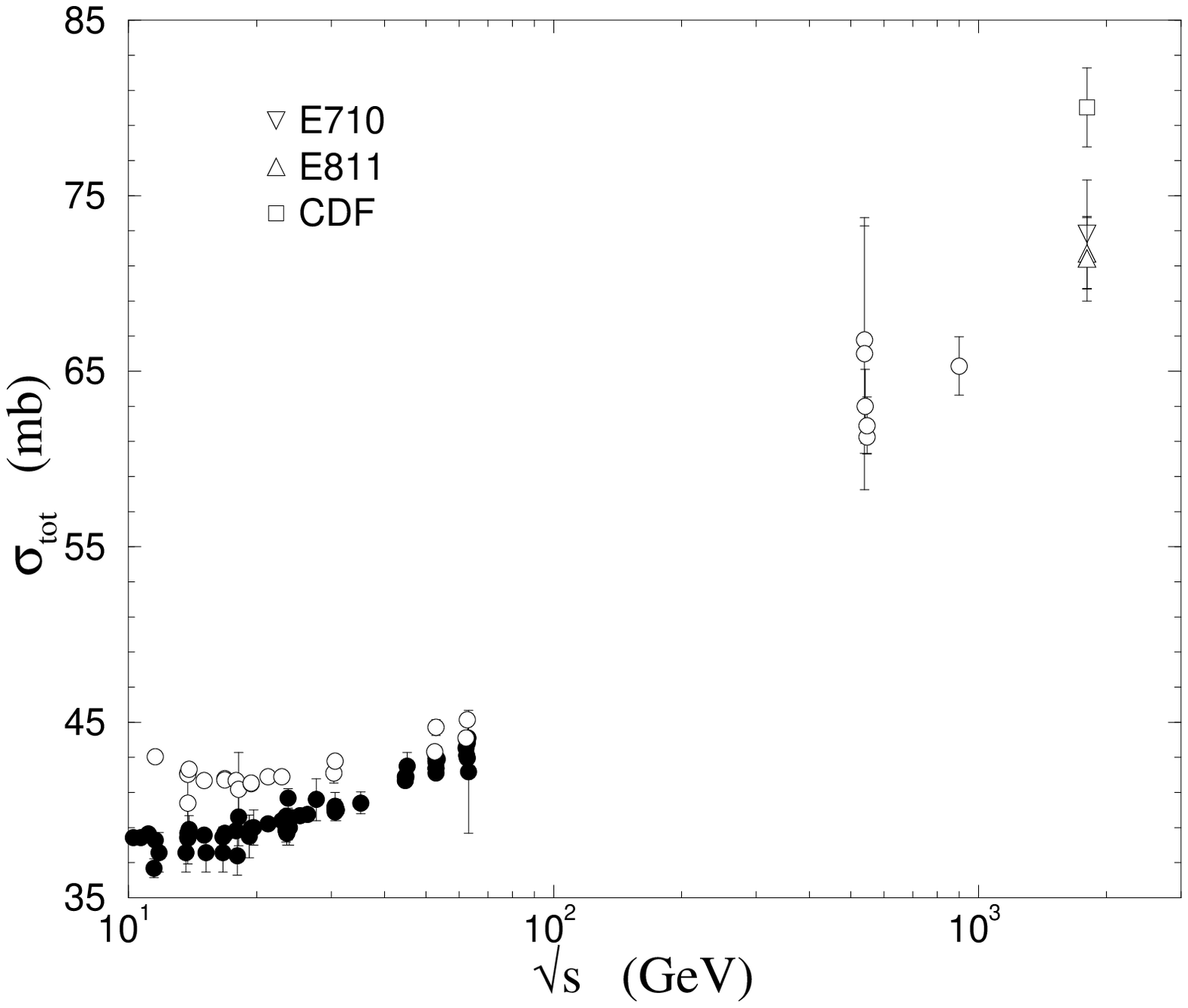}
\vglue -7.1cm
\hglue 8.3cm
\includegraphics[height=.30\textheight]{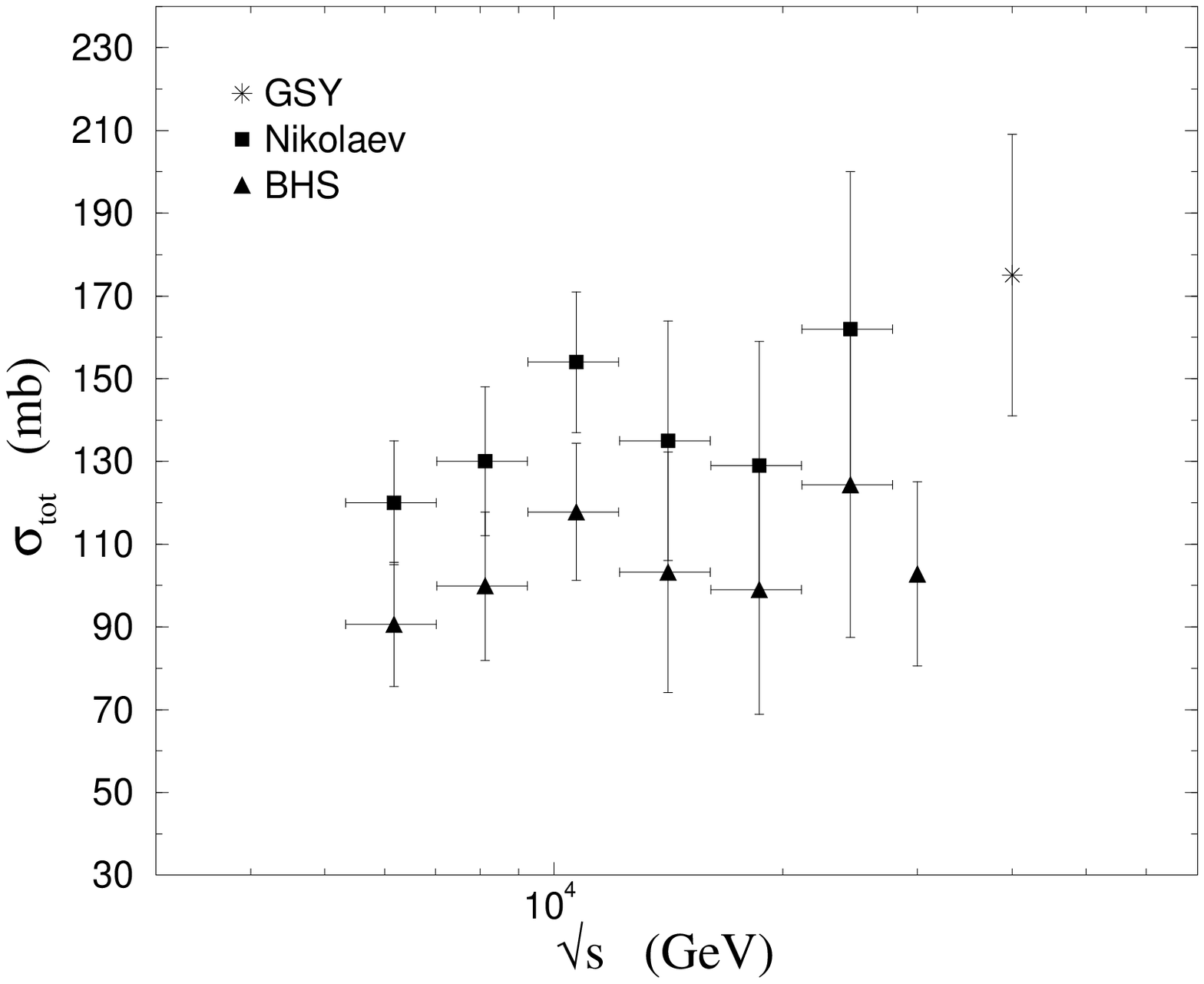}
\caption{The $pp$ (black symbols) and $\bar{p}p$ (white symbols) total cross 
section above 
$\sqrt{s} = 10$ GeV from accelerator experiments (left) and estimations
of $pp$ total cross section  from cosmic ray experiments (right)
by Gaisser, Sukhatme, and Yodh (GSY) \cite{gsy}, Nikolaev \cite{niko},
and Block, Halzen, and Stanev (BHS) \cite{bhs,private}.}
\end{figure*}

Based on these facts, we consider important at the moment to investigate these
experimental discrepancies and examine its consequences
in terms of extrema bounds for the Pomeron intercept. 

Since recent analysis showed that the parameters of Regge
fits are stable for a cutoff $\sqrt s \sim 9$ GeV \cite{cudell},
in what follows we consider experimental data on $\sigma_{tot}$ and $\rho$ above
$\sqrt s = 10$ GeV. We use the data sets compiled
and analyzed by the Particle Data Group \cite{pdg}, to which we add the new E811 data on 
$\sigma_{tot}$ and $\rho$ 
at $1.8$ TeV \cite{E811}. The statistic and systematic errors have been added in 
quadrature.

In order to investigate the effects of the discrepancies in a quantitative way, 
we select different ensembles
of $\sigma_{tot}$ data that include all the $\rho$ results above $10$ GeV. First we only consider 
accelerator data in two ensembles with the following notation:
Ensemble \textbf{I}: $\sigma^{pp}_{tot}$ and $\sigma^{\bar{p}p}_{tot}$ data ($10 \le \sqrt{s} \le 
900$ GeV) + CDF datum ($\sqrt{s} = 1.8$ TeV);
ensemble \textbf{II}: $\sigma^{pp}_{tot}$ and $\sigma^{\bar{p}p}_{tot}$ data ($10 \le \sqrt{s} \le 
900$ GeV) + E710/E811 data ($\sqrt{s} = 1.8$ TeV). 
Ensemble I represents the faster increase scenario for the rise of $\sigma_{tot}$ from
accelerator data and ensemble II the slowest one. These ensembles are then combined with
the highest and lowest estimations for $\sigma_{tot}^{pp}$ from cosmic-ray experiments,
namely, the Nikolaev and Gaisser, Sukhatme, and Yodh (NGSY) results and the Block, Halzen, 
and Stanev (BHS) results, respectively. These new ensembles are denoted by 
Ensemble \textbf{I + NGSY} and Ensemble \textbf{II + BHS}.

As in the previous paper, we consider both individual fits to $\sigma_{tot}$, and 
simultaneous fits to $\sigma_{tot}$ and $\rho$, either in the case where the subtraction 
constant is considered as a free fit parameter or assuming $K = 0$ in Eq. (3). The fits have
been performed with the program CERN-MINUIT and the errors in the fit parameters
correspond to an increase of the $\chi^2$ by one unit.

In the case of accelerator data only the fit results  
for $\sigma_{tot}$ and $\rho$ with ensembles I and II
are displayed in Table I and Figs. 2, 3 and 4. The results concerning the combination
of these ensembles
with the estimations from cosmic-ray experiments, namely, ensembles I + NGSY and
II + BHS, are shown in Table II and Figs. 5, 6 and 7. In the last two cases we present the
curves and experimental information in the region from $500$ GeV to $50$ TeV, since
the results are the same at lower energies, as can be seen in the corresponding
results for $\rho(s)$.

\begin{table*}
\caption{Individual and global fits to $\sigma_{tot}$ and $\rho$ with 
Ensemble I (CDF datum) and Ensemble II (E710/E811 data) and the subtraction
constant $K = 0$ or as a free fit parameter.}
\begin{ruledtabular}
\begin{tabular}{ccccccc}
Fit: & \multicolumn{2}{c}{Individual - $\sigma_{tot}$} 
& \multicolumn{2}{c}{Global - $\sigma_{tot}$ and
$\rho$ with $K=0$} & \multicolumn{2}{c}{Global - $\sigma_{tot}$ and $\rho$ with $K$ free}\\
Ensemble: &  I &   II &  I &  II &  I &  II \\ 
\hline
$\epsilon$ & 0.096$\pm$0.005 & 0.085$\pm$0.004 & 0.098$\pm$0.004 & 0.090$\pm$0.003 & 
0.095$\pm$0.005 & 0.085$\pm$0.003 \\
$X$ (mb) & 18$\pm$1 & 20$\pm$1 & 18$\pm$1 & 19$\pm$1 & 19$\pm$1 &  21$\pm$1\\
$\eta_{+}$ & 0.31$\pm$0.04 & 0.38$\pm$0.04 & 0.32$\pm$0.02 & 0.35$\pm$0.02 & 0.35$\pm$0.04 & 
0.41$\pm$0.04 \\
$Y_{+}$ (mb) & 55$\pm$5 & 62$\pm$8 & 56$\pm$3 & 58$\pm$3 & 62$\pm$7 &  71$\pm$8\\
$\eta_{-}$ & 0.42$\pm$0.04 & 0.42$\pm$0.04 & 0.53$\pm$0.02 & 0.53$\pm$0.02 & 0.52$\pm$0.02 &  
0.52$\pm$0.02\\
$Y_{-}$ (mb) & -17$\pm$4 & -17$\pm$4 & -30$\pm$4 & -30$\pm$4 & -29$\pm$4 & -29$\pm$4 \\
$K$ & - & - & 0 & 0 & 74$\pm$61 & 136$\pm$64 \\
$\textrm{No. DOF}$ & 87 & 89 & 147 & 149 & 146 & 148 \\
$\chi^2 /\textrm{DOF}$ & 0.95 & 0.94 & 1.08 & 1.10 & 1.07 & 1.07 \\
\end{tabular}
\end{ruledtabular}
\end{table*}

\begin{figure*}
\vspace{1.0cm}
\vglue -1.0cm
\hglue -8.5cm
\includegraphics[height=.30\textheight]{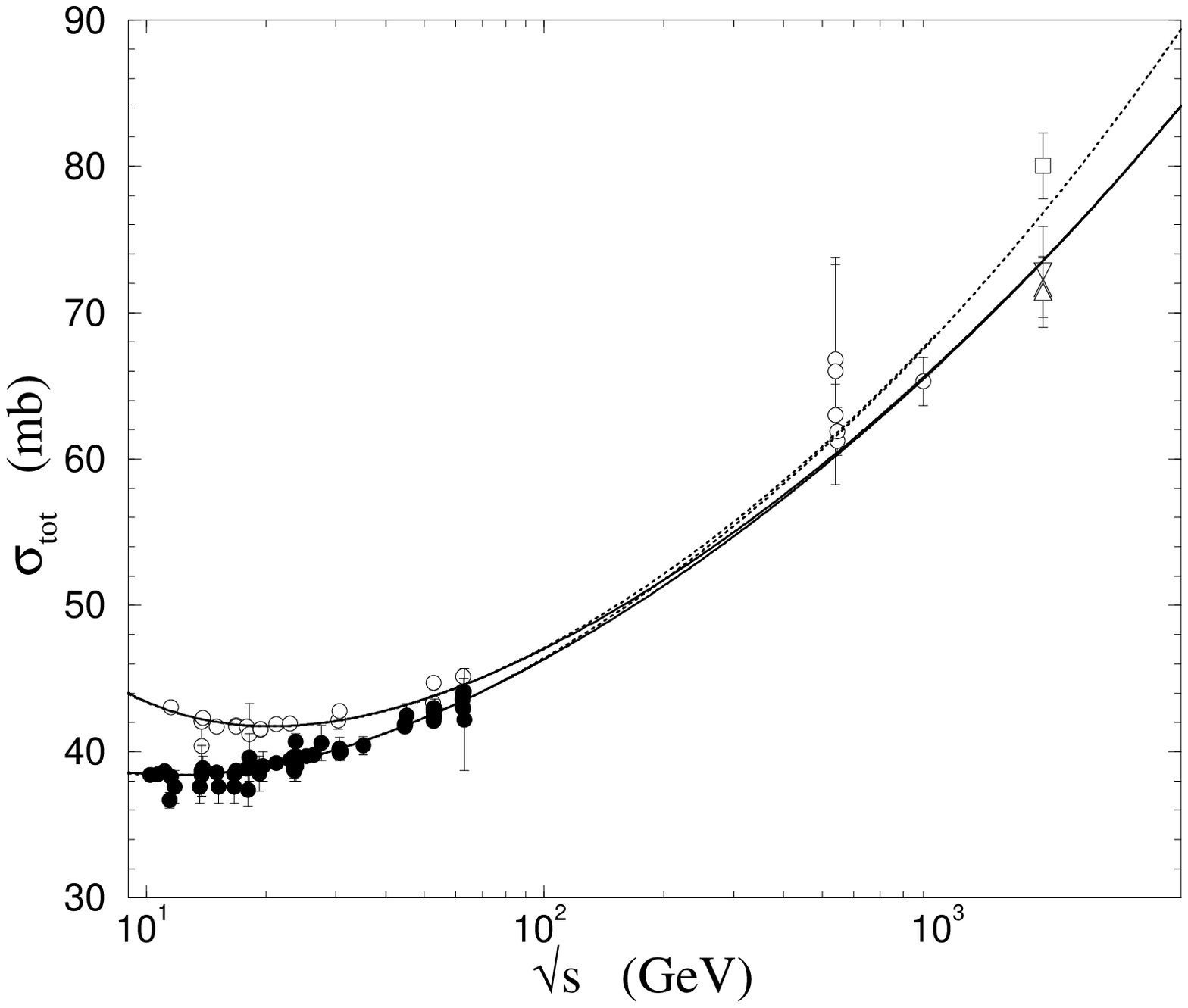}
\vglue -7.1cm
\hglue 8.3cm
\includegraphics[height=.30\textheight]{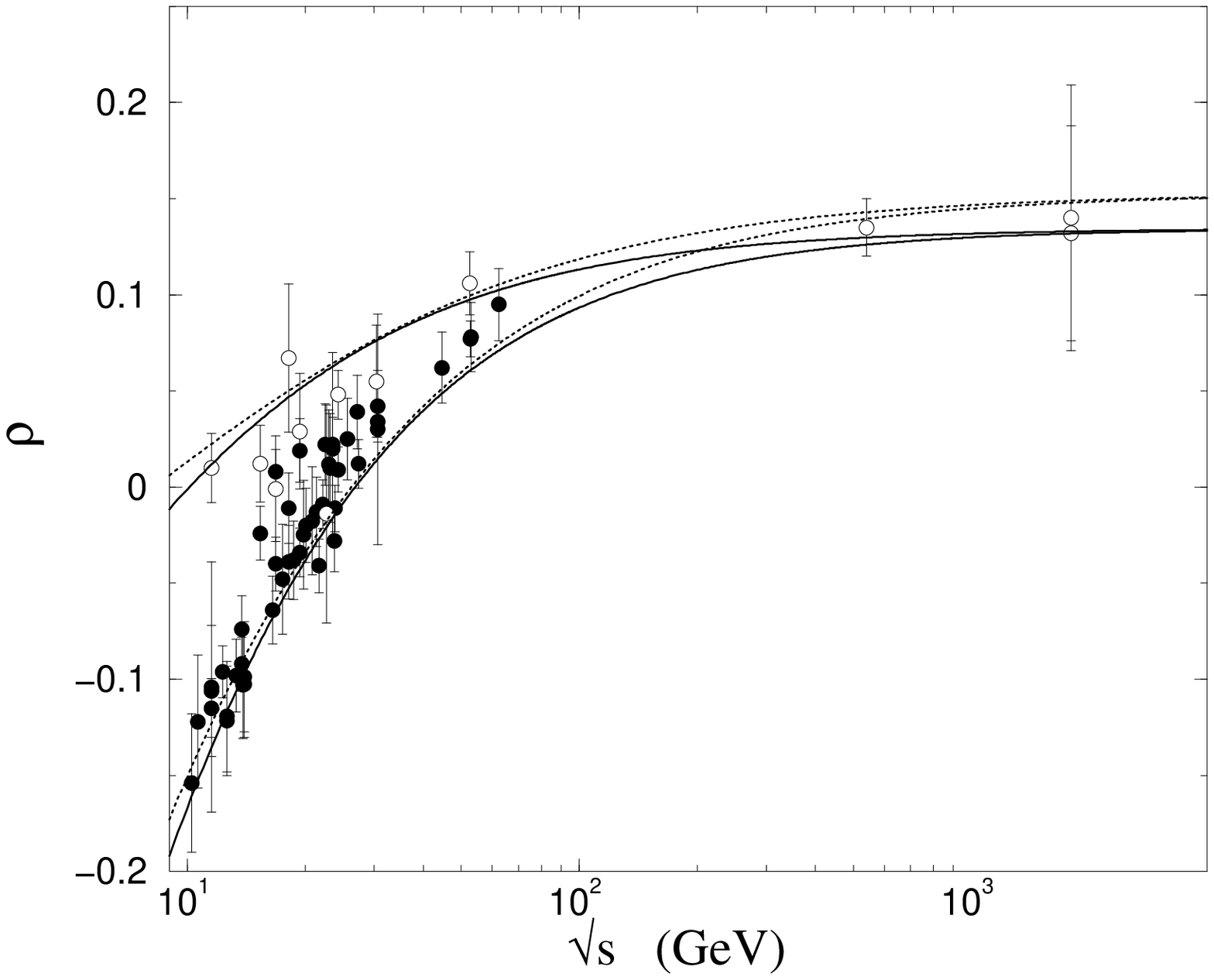}
\caption{Fits to $pp$ (black symbols) and $\bar{p}p$ (white symbols) total cross section 
data from ensembles I (dotted curves) and II (solid curves) and the corresponding 
predictions for $\rho(s)$ with $K = 0$.}
\end{figure*}

\begin{figure*}
\vspace{1.0cm}
\vglue -1.0cm
\hglue -8.5cm
\includegraphics[height=.30\textheight]{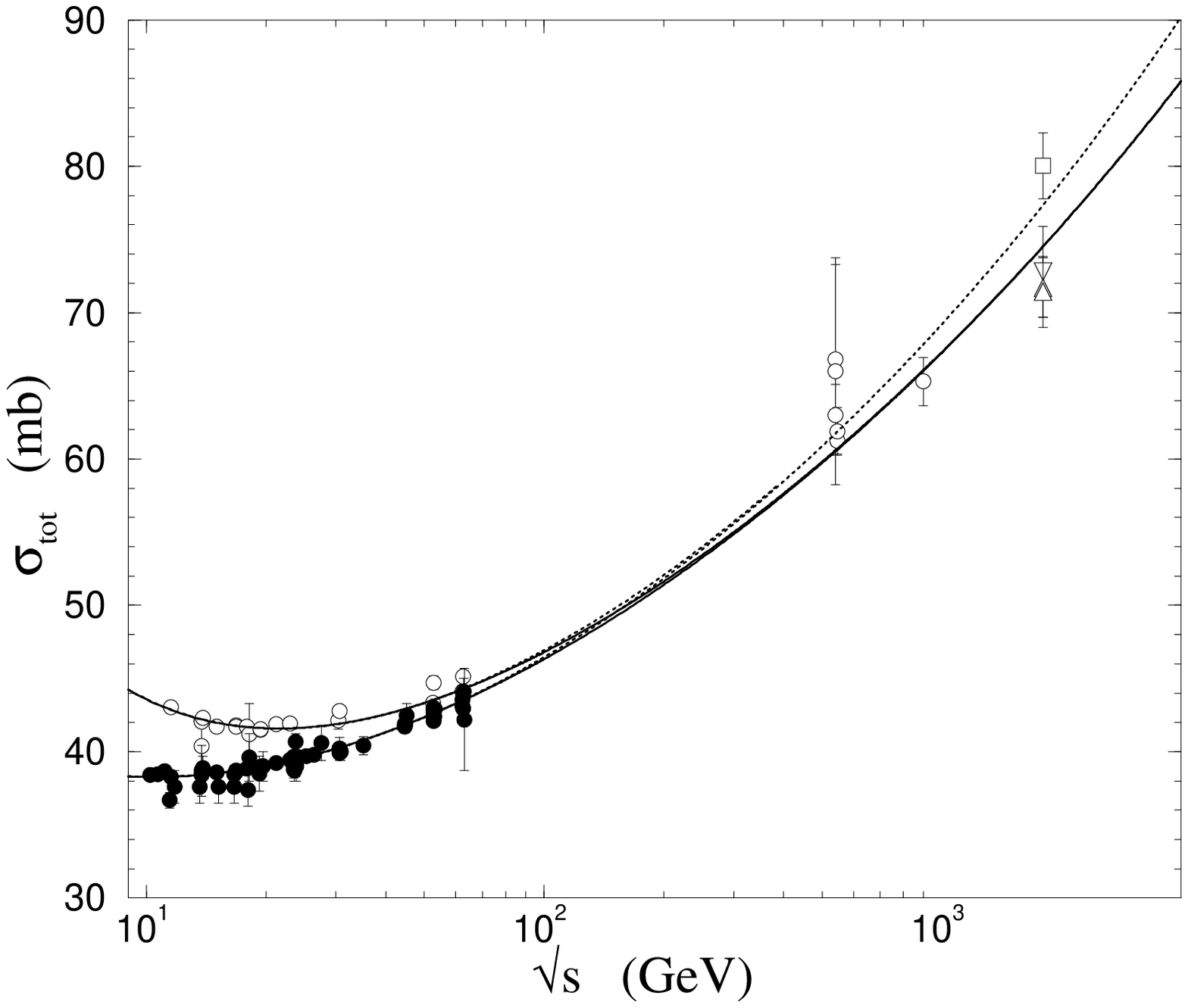}
\vglue -7.1cm
\hglue 8.3cm
\includegraphics[height=.30\textheight]{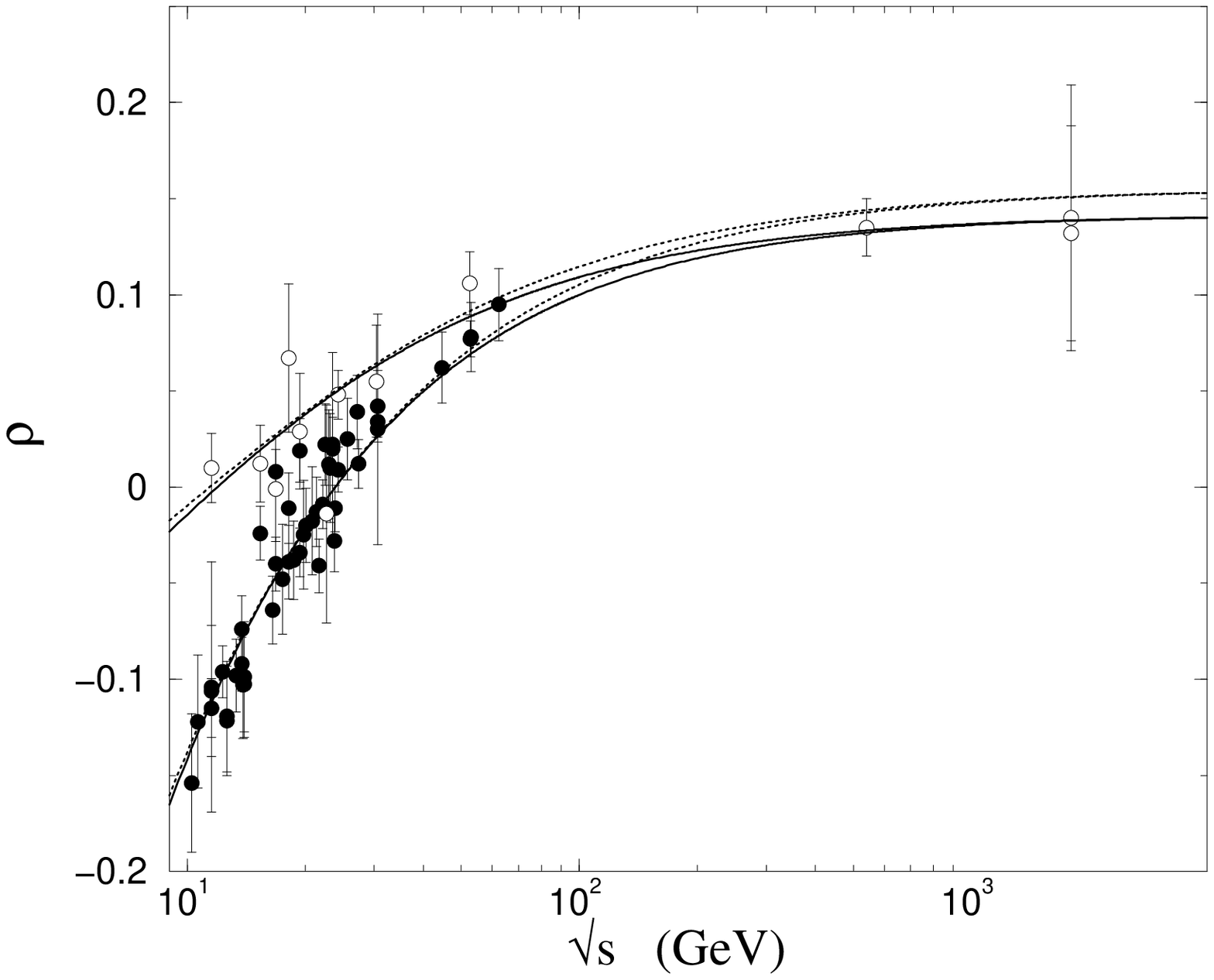}
\caption{Simultaneous fits to $\sigma_{tot}(s)$ and $\rho(s)$ data
from ensembles I
(dotted curves) and II 
(solid curves), with $K = 0$.}
\end{figure*}

\begin{figure*}
\vspace{1.0cm}
\vglue -1.0cm
\hglue -8.5cm
\includegraphics[height=.30\textheight]{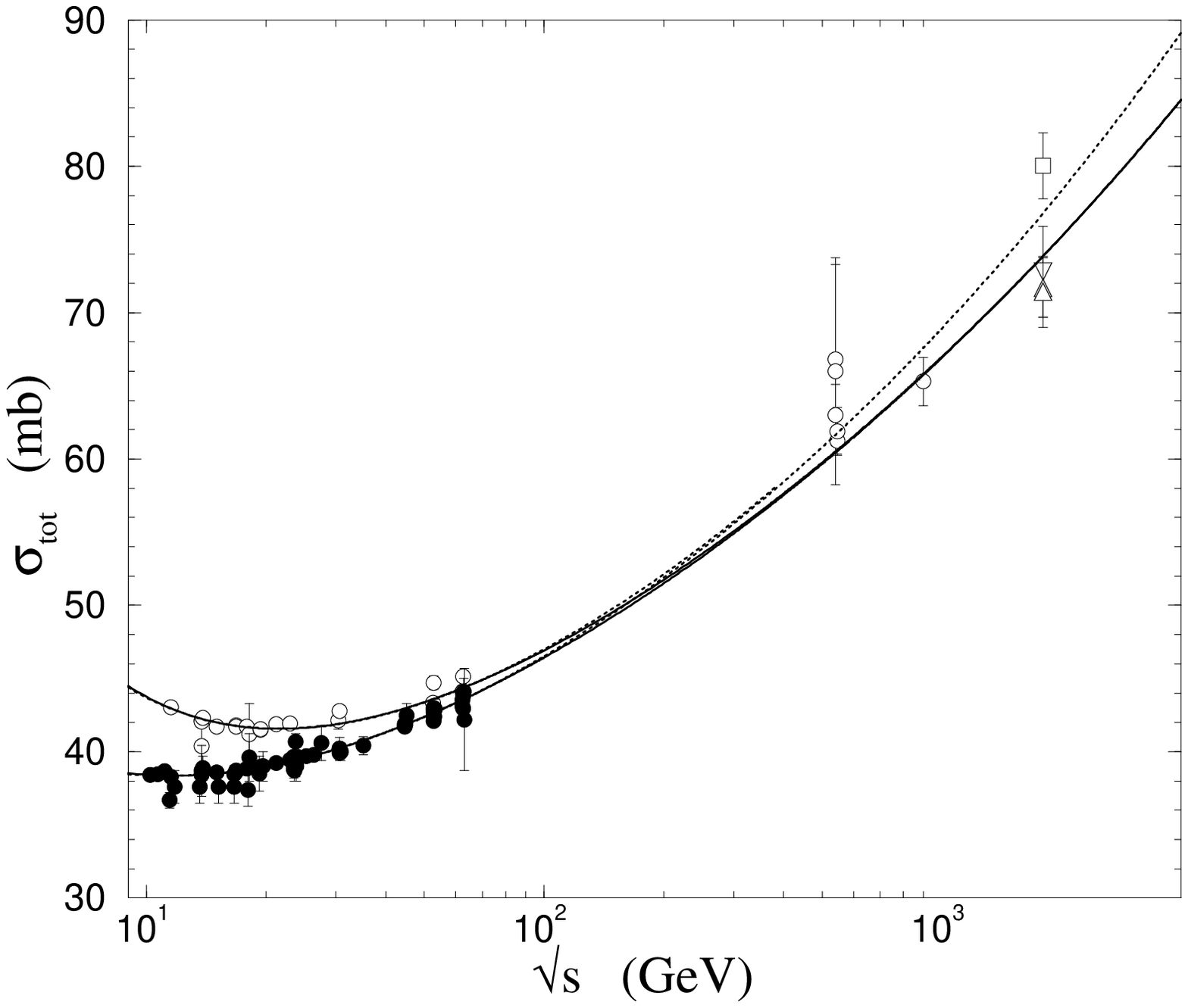}
\vglue -7.1cm
\hglue 8.3cm
\includegraphics[height=.30\textheight]{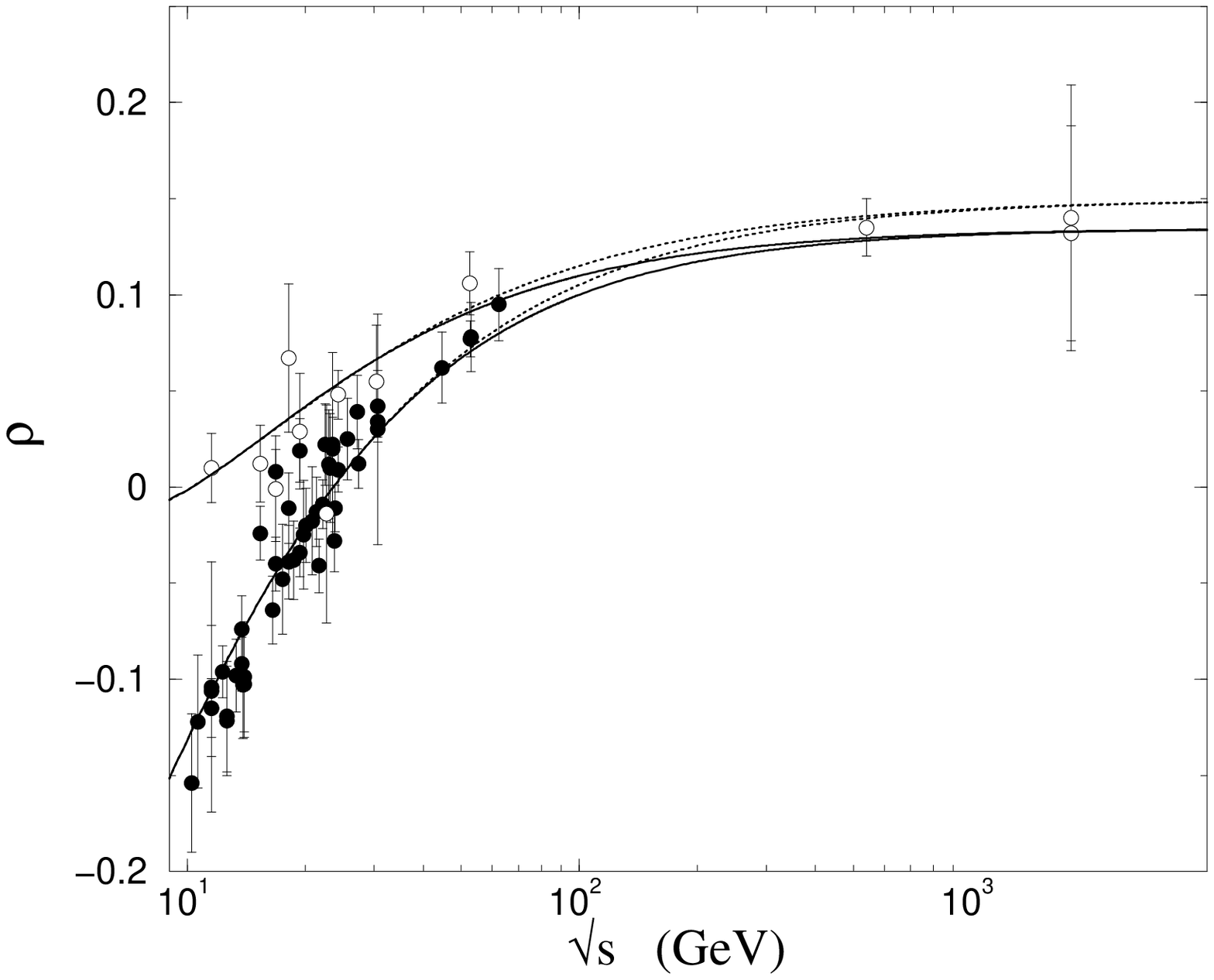}
\caption{Simultaneous fits to $\sigma_{tot}(s)$ and $\rho(s)$ data
from ensembles I (dotted curves) and II  (solid curves), with $K$
as a free fit parameter.}
\end{figure*}

\begin{table*}
\caption{Individual and global fits to $\sigma_{tot}$ and $\rho$ with
Ensemble I + NGSY and Ensemble II + BHS  and the subtraction
constant $K = 0$ or as a free fit parameter.}
\begin{ruledtabular}
\begin{tabular}{ccccccc}
Fit: & \multicolumn{2}{c}{Individual - $\sigma_{tot}$}
& \multicolumn{2}{c}{Global - $\sigma_{tot}$ and
$\rho$ with $K=0$} & \multicolumn{2}{c}{Global - $\sigma_{tot}$ and $\rho$
with $K$ free}\\
Ensemble: & I + NGSY & II + BHS & I + NGSY & II + BHS & I + NGSY & II +
BHS \\
\hline
$\epsilon$ & 0.104$\pm$0.005 & 0.085$\pm$0.003 & 0.102$\pm$0.004 &
0.089$\pm$0.003 & 0.100$\pm$0.004 & 0.085$\pm$0.003 \\
$X$ (mb) & 16$\pm$1 & 20$\pm$1 & 17$\pm$1 & 19$\pm$1 & 17$\pm$1 &
21$\pm$1\\
$\eta_{+}$ & 0.28$\pm$0.03 & 0.38$\pm$0.04 & 0.30$\pm$0.02 & 0.35$\pm$0.02
& 0.32$\pm$0.03 &
0.41$\pm$0.04 \\
$Y_{+}$ (mb) & 51$\pm$4 & 62$\pm$7 & 55$\pm$3 & 58$\pm$3 & 58$\pm$5 &
71$\pm$9\\
$\eta_{-}$ & 0.42$\pm$0.04 & 0.42$\pm$0.04 & 0.52$\pm$0.02 & 0.53$\pm$0.02
& 0.52$\pm$0.02 & 0.52$\pm$0.03\\
$Y_{-}$ (mb) & -17$\pm$4 & -17$\pm$4 & -29$\pm$4 & -30$\pm$4 & -29$\pm$4 &
-29$\pm$4 \\
$K$ & - & - & 0 & 0 & 41$\pm$52 & 135$\pm$68 \\
$\textrm{No. DOF}$ & 94 & 96 & 154 & 156 & 153 & 155 \\
$\chi^2 /\textrm{DOF}$ & 1.01 & 0.89 & 1.11 & 1.06 & 1.11 & 1.03 \\
\end{tabular}
\end{ruledtabular}
\end{table*}

\begin{figure*}
\vspace{1.0cm}
\vglue -1.0cm
\hglue -8.5cm
\includegraphics[height=.30\textheight]{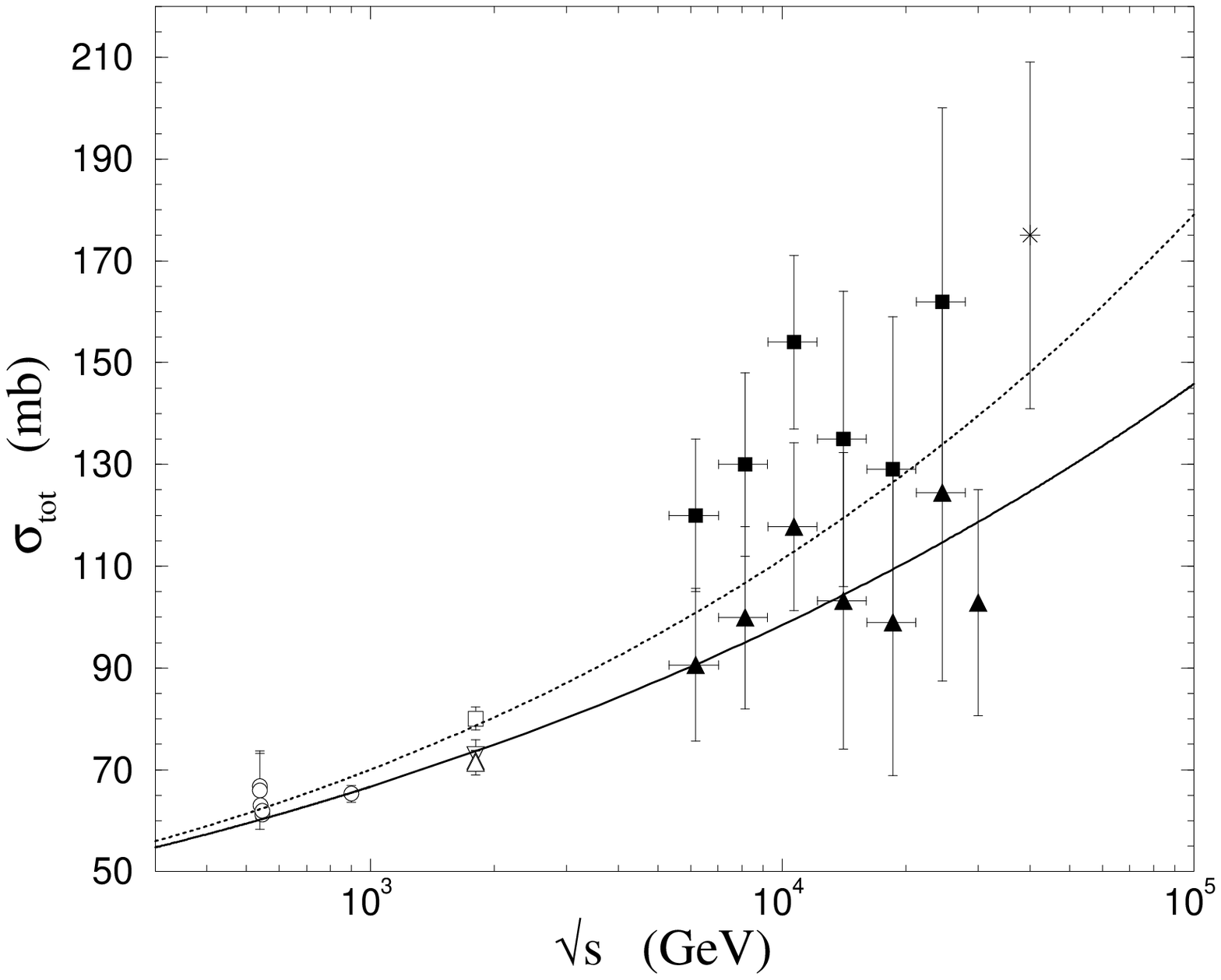}
\vglue -7.1cm
\hglue 8.3cm
\includegraphics[height=.30\textheight]{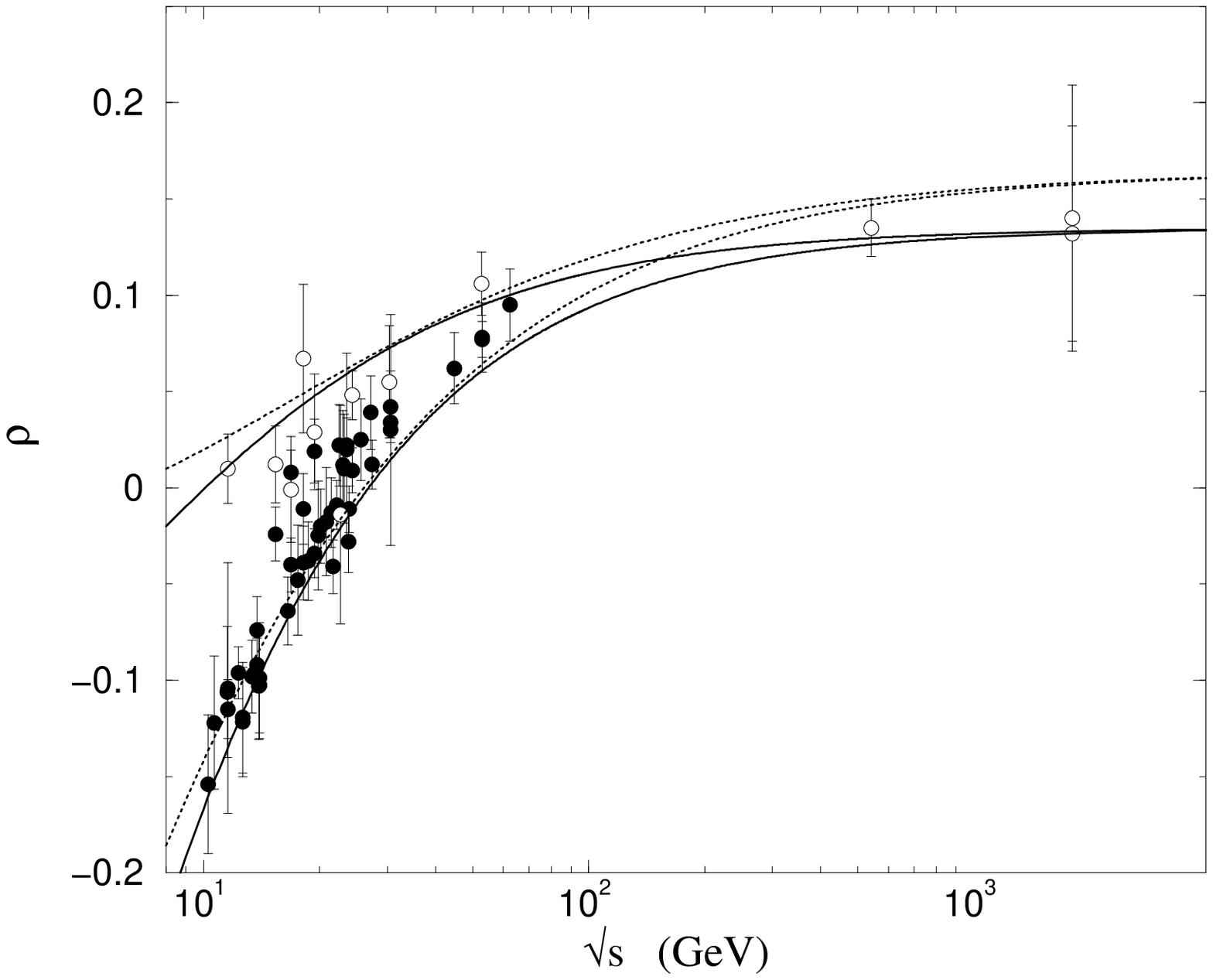}
\caption{Fits to $pp$ and $\bar{p}p$ total cross section data from ensembles 
I + NGSY
(dotted curves) and II + BHS 
(solid curves) and the corresponding 
predictions for $\rho(s)$ with $K = 0$.}
\end{figure*}

\begin{figure*}
\vspace{1.0cm}
\vglue -1.0cm
\hglue -8.5cm
\includegraphics[height=.30\textheight]{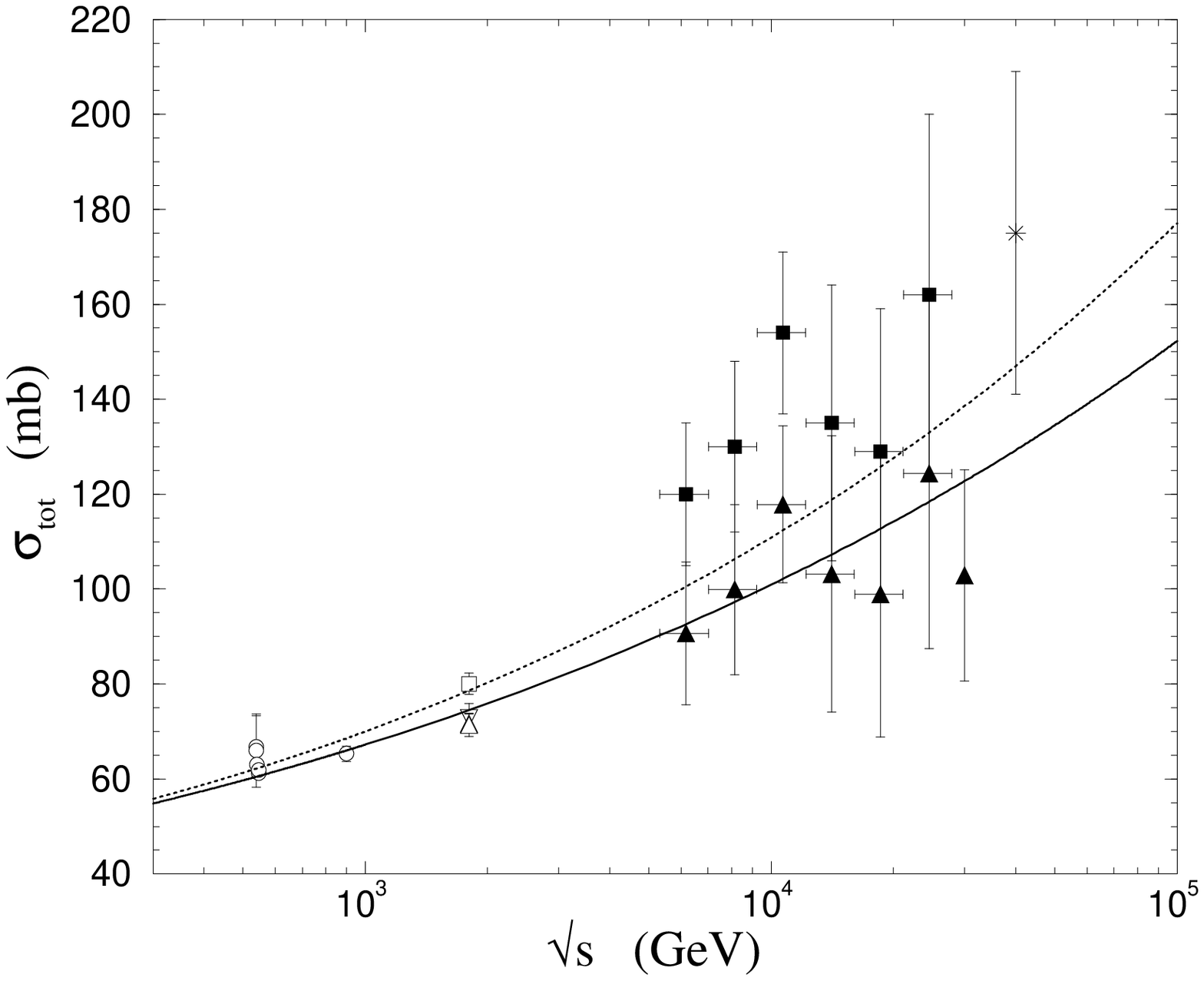}
\vglue -7.1cm
\hglue 8.3cm
\includegraphics[height=.30\textheight]{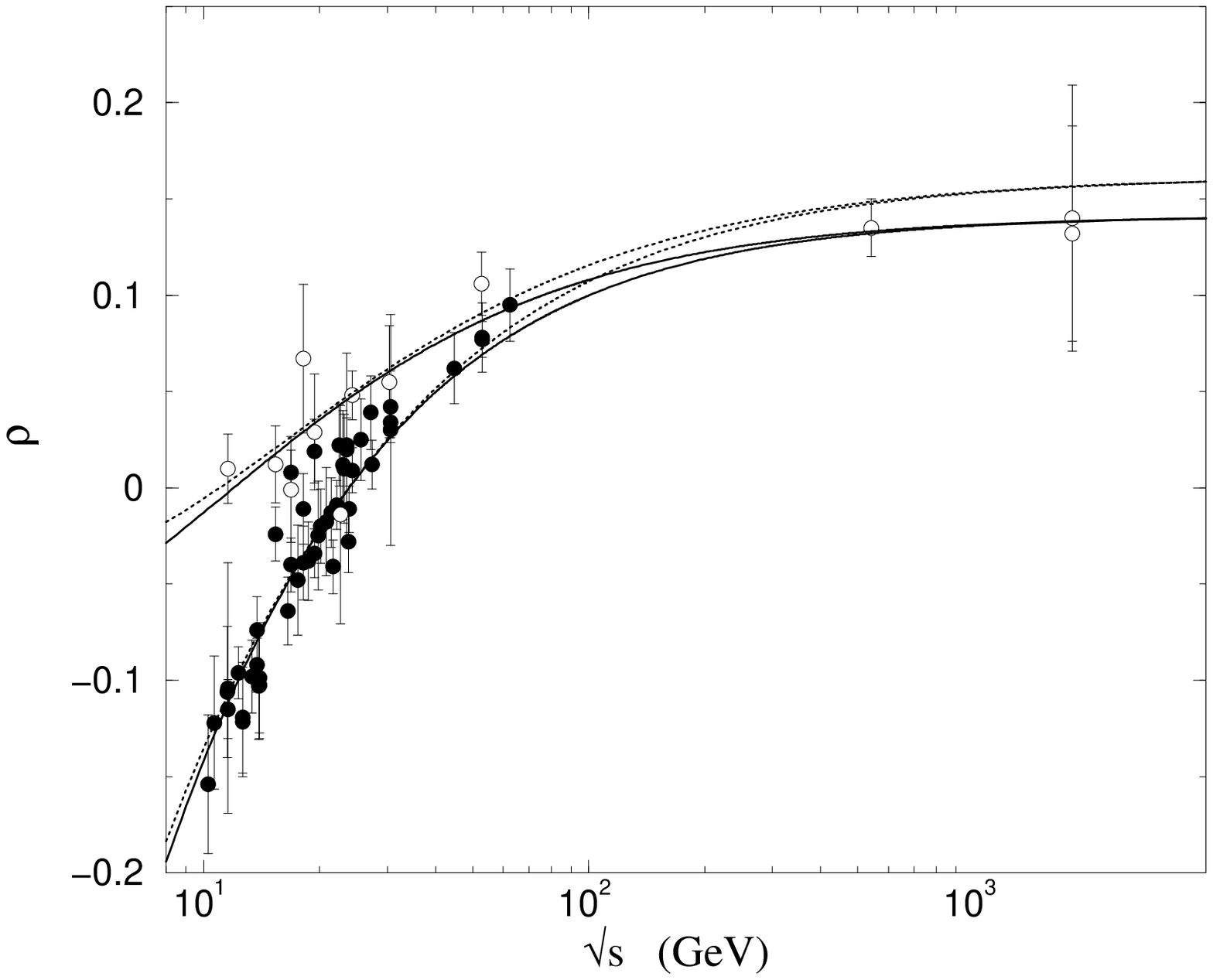}
\caption{Simultaneous fits to $\sigma_{tot}(s)$ and $\rho(s)$ data
from ensembles I + NGSY
(dotted curves) and II + BHS 
(solid curves), with $K = 0$}
\end{figure*}

\begin{figure*}
\vspace{1.0cm}
\vglue -1.0cm
\hglue -8.5cm
\includegraphics[height=.30\textheight]{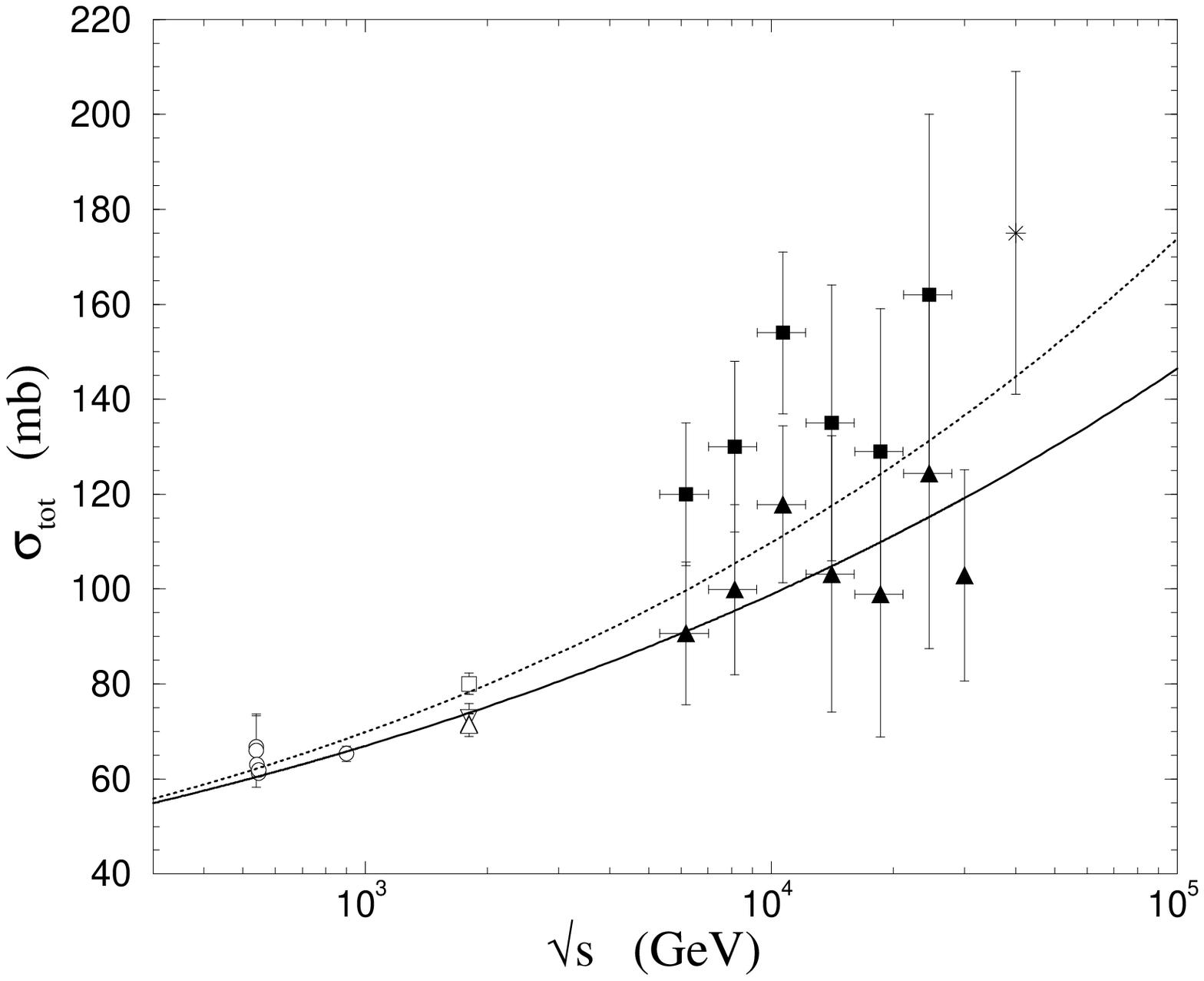}
\vglue -7.1cm
\hglue 8.3cm
\includegraphics[height=.30\textheight]{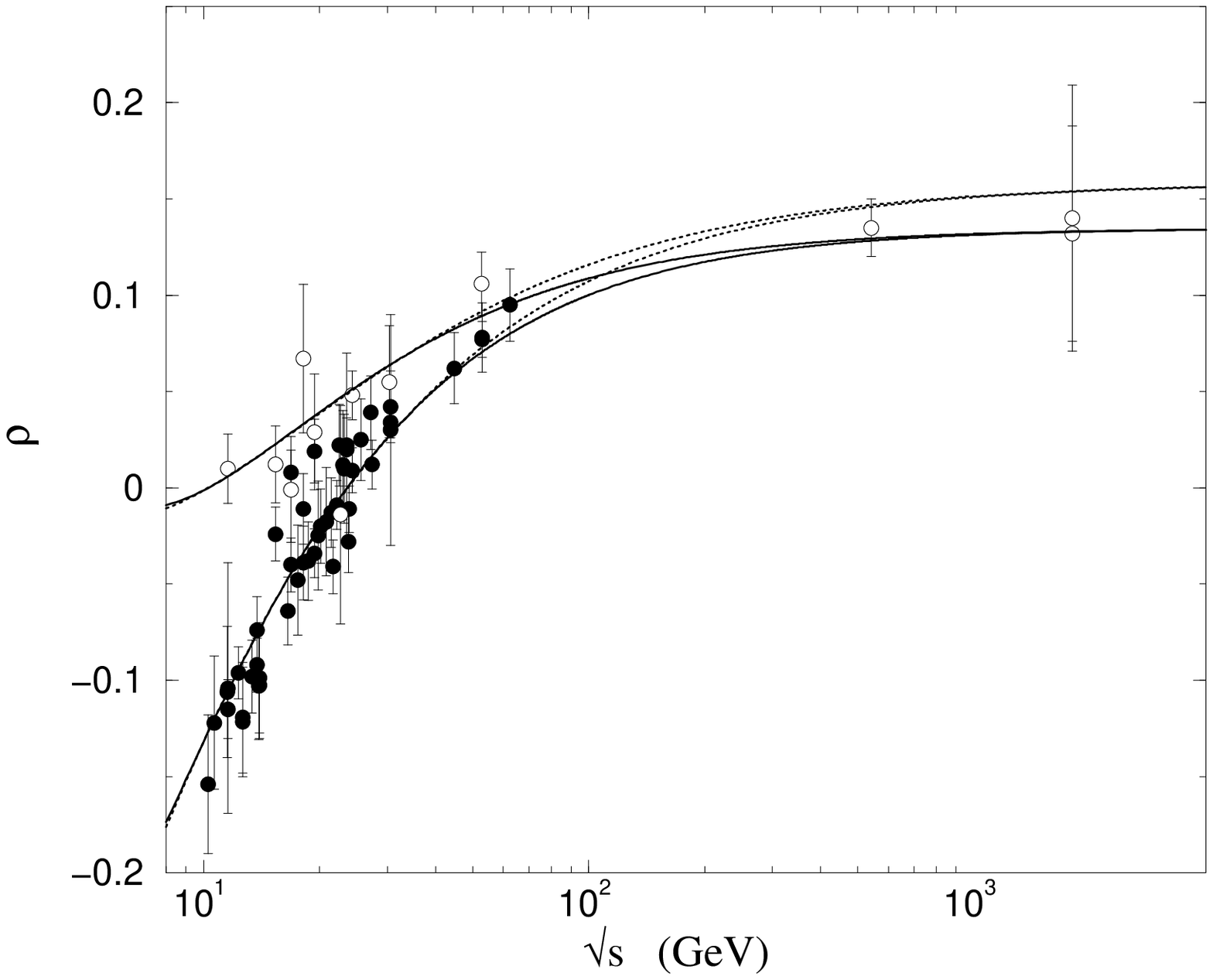}
\caption{Simultaneous fits to $\sigma_{tot}(s)$ and $\rho(s)$ data
from ensembles I + NGSY
(dotted curves for $pp$ and dashed for $\bar{p}p$) and II + BHS
(solid curves for $pp$ and dot dashed for $\bar{p}p$), with $K$
as a free fit parameter}
\end{figure*}

\vfill\eject
\newpage

\section{Conclusions and final remarks}

In this analysis we have used the experimental information presently available in the
accelerator domain, including the recent E811 results on $\sigma_{tot}^{\bar{p}p}$
and $\rho$ at $1.8$ TeV, and also the highest and lowest estimations for $\sigma_{tot}^{pp}$
from cosmic-ray experiments.

From Table I (only accelerator data), we may infer the following upper and lower
values for the Pomeron intercept: $\alpha^{upper}_{\tt I\!P}(0)=1.098\pm 0.004$
(global fits to ensemble I, with $K = 0$)
and $\alpha^{lower}_{\tt I\!P}(0)=1.085\pm 0.004$
(individual fit to $\sigma_{tot}$ from ensemble II), 
with bounds $1.102$ and $1.081$, respectively. 

Adding the cosmic-ray information, Table II, we infer
$\alpha^{upper}_{\tt I\!P}(0)=1.104\pm 0.005$
(individual fit to $\sigma_{tot}$ from ensemble I + NGSY)
and $\alpha^{lower}_{\tt I\!P}(0)=1.085\pm 0.003$
(global fits to ensemble II + BHS, and $K$ as a free fit parameter
or individual fit to $\sigma_{tot}$ from this ensemble), 
with bounds $1.109$ and $1.082$,
respectively.

Our approach, and the above values and bounds, may be
compared with some representative results obtained by other authors,
which are displayed in Table III and are
reviewed in what follows. The fits by Donnachie and Landshoff (DL) \cite{dl92}
have been performed to only $pp$ and $\bar{p}p$ total cross section data, above $10 $ GeV and 
with the E710 result at $1.8$ TeV. The CDF Collaboration (CDF), based on their
result for the $\sigma_{tot}^{\bar{p}p}$, obtained a higher value for the intercept
\cite{CDF}.
Further analyses, through the extended Regge parametrization, included also the
E811 result. In the work by Cudell, Kang, and Kim (CKK) \cite{ckk} only
$pp$ and $\bar{p}p$ data above $10$ GeV have been fitted. The analysis by
Covolan, Montanha and Goulianos (CMG) \cite{cmg} (using both a Born level and Eikonal 
parametrizations) involved global fits to $pp$, $\bar{p}p$, $\pi^{\pm} p$ and
$k^{\pm} p$ at $\sqrt s \geq 6$ GeV.
The COMPETE Collaboration (COMPETE) \cite{cudell} treated simultaneous fits to 
$\sigma_{tot}$ and $\rho$ in
global fits
to $pp$, $\bar{p}p$, meson-$p$, $\gamma p$ and $\gamma \gamma$ above $9$ GeV.
All these results concerned only
accelerator data. In Ref. \cite{alm03}, \'Avila, Luna and Menon (ALM) included also
some cosmic-ray estimations for the $\sigma_{tot}^{pp}$ and made use of the 
original DL parametrization.
It is also shown in Table III a recent theoretical result by R. A. Janik 
(Janik) \cite{janik} through
a nonperturbative approach and using the AdS/CFT correspondence.

\begin{table}
\caption{Some representative values, bounds and limits for the soft Pomeron
intercept and those obtained in this work.}
\begin{ruledtabular}
\begin{tabular}{ccc}
%& \multicolumn{2}{c}{Ensemble I}&\multicolumn{2}{c}{Ensemble II}\\
 & $\epsilon = \alpha_{\tt I\!P}(0) - 1$ & bounds / limits\\
\hline
 DL \cite{dl92} & $0.0808$ & - \\
 CDF \cite{CDF} & $0.112 \pm 0.013$ & - \\
 CKK \cite{ckk} & $0.096^{+ 0.012}_{-0.009}$ & - \\
 CMG \cite{cmg} & $0.104 \pm 0.002$ (Born) & - \\
  & $0.122 \pm 0.002$ (Eikonal) & - \\
COMPETE \cite{cudell} & $0.093 \pm 0.002$ & - \\
 ALM \cite{alm03} & - &  $0.0790 - 0.0940$\\
 Janik \cite{janik} &  - & $0.0729 - 0.083$ \\
This work & $0.085 \pm 0.004$ (lower) & $0.081$ \\
          & $0.104 \pm 0.005$ (upper) & $0.109$ \\
\end{tabular}
\end{ruledtabular}
\end{table}

In this work we have presented all the possible fits to
$pp$ and $\bar{p}p$ data, above $10$ GeV, through the extended Regge
parametrization, exploring the contrasting data and
 the faster and the slower increase scenarios for
the rise of the total cross section, allowed by the experimental information
presently available.

From Table III, our results exclude the values for the Pomeron intercept obtained by
CMG (in the case of the eikonal parametrization), the lower
bounds by ALM and Janik and the mean value by the CDF Collaboration.
The DL result is barely compatible with our lower limit.
It should be noted that, if the same ensemble is fitted, the introduction
of nondegenerate trajectories results in a slightly increase of the Pomeron
intercept. For example, fit to all the accelerator data above $10$ GeV
(including the CDF and the E710/E811 values), leads to
$\epsilon = 0.086 \pm 0.003$ and
$\epsilon = 0.089 \pm 0.004$, in the cases of degenerate (DL) and
nondegenerate parametrizations, respectively. 

From figures (2) to (7), we see that in all the cases investigated
the $\rho$ parameter is better described with ensembles I and
I + BHS, a result that is also roughly supported by the
$\chi^2$/DOF (Tables I and II). We understand that this picture favors the
E710/E811 results. This conclusion is contrary to that 
obtained by CMG \cite{cmg} and more
recently  by the COMPETE Collaboration \cite{compete}.

As a next step it may be important to investigate the consequences of the
above extrema bounds in fittings to
$p$-mesons, $p \gamma$, and $\gamma \gamma$  scattering,
with focus in the ratio of strengths of the Pomeron
exchange (quark counting and factorization).

\begin{acknowledgments}
We are thankful to FAPESP for financial support (Contracts No. 00/00991-7, and 
No. 00/04422-7) and to P. Valin, J. Montanha Neto, and A. F. Martini for discussions.
We are also grateful to M. M. Block for sending us the total cross section results
concerning the BHS analysis.
\end{acknowledgments}

\end{document}